\documentclass[12pt,a4paper]{article}

\usepackage{amsmath}
\usepackage{amsthm}
\usepackage{amscd}
\usepackage{amsxtra}
\usepackage{upref}
\usepackage{amsfonts}
\usepackage{amssymb}
\usepackage{graphicx}
\usepackage{url}
\usepackage{cite}
\usepackage{pifont}

\def\beq{\begin{equation}}
\def\eeq{\end{equation}}
\def\bea{\begin{eqnarray}}
\def\eea{\end{eqnarray}}

\theoremstyle{definition}

\numberwithin{equation}{section}

\title{Antioscillons}
\author{L. Mersini-Houghton\\
Department of Physics and Astronomy, UNC Chapel Hill, NC 27599, USA.}
\begin{document}
\maketitle
\date{\today}
\begin{abstract}
This proceeding is based on a talk I gave at the 13th Marcel Grossmann Meeting in Stockholm Sweden, on the role of topology on bubble collisions. 
\end{abstract}
\section{Introduction }
In the framework of eternal inflation bubble universes undergo collisions.
At finite temperatures bubbles have nontrivial topology. The $0(4)$ symmetry characteristic of Coleman-Deluccia bubbles at $T=0$ is broken by finite temperature effects to solutions that are $O(3)$ symmetric in the spatial directions and periodic in the Euclidean time direction
with period $\tau = it = \beta = T^{-1}$, as shown in ~\cite{ref:Linde}.
For $T > R_0^{-1}$ (where $R_0$ is the initial nucleation radius of the bubble), individual bubbles overlap in the $\tau$ direction, and the resulting bubbles acquire a cylindrical topology in Euclidean space. The periodicity with respect to the temperature leads to bubbles of topology $R^3 x S^1$ where $R^3$ describes the 3-dimensional spatial direction and $S^1$ the periodic ``time'' dependence.
As shown in~\cite{ref:Linde}, the action $S_4(\phi)$ of $O(4)$ bubbles at $T=0$ is therefore replaced by $S_3(\phi)/T$ for $O(3)$ bubbles with cylindrical symmetry, periodic in $\tau$.
The change in the topology of the bubbles induced by finite temperature effects leads these topologies to admit two types of field configurations, the twisted fields $\psi$ and untwisted field $\phi$, satisfying the following boundary conditions
\begin{equation}
  \psi(x,t) = -\psi(x,t+i\beta)
  \phi(x,t) = \phi(x,t+i\beta)  \, .
\end{equation}
Nontrivial topologies of bubbles are common, even at $T=0$.
For example, it was recently shown~\cite{ref:bousso} that the common shape for a cluster of bubbles with two or more collisions is a torus shape. A torus topology also admits the twisted and untwisted configurations.
An interesting aspect of the field configurations is the emergence of novel structures of energy they give rise to during bubble collisions. One of these structures, the oscillon, was discovered 2 decades ago and studied extensively in the context of phase transitions.~\cite{ref:Gleiser}. The basic mechanism that gives rise to such structures is a parametric amplification of fluctuations of a field around its vacuum state, 
such that it is driven to large fluctuations near the inflection point of the potential.
The large fluctuations condense into quasistable Gaussian oscillating lumps of energy, thus the name oscillons. They can be induced by a variety of mechanisms, 
such as a sudden quench of the potential (relative to the fields relaxation time),
by thermal noise,
or by nonlinear couplings to itself and other fields.
I describe here how fluctuations in the twisted field in the bubble produce structures, the antioscillons, from collisions by the same mechanism that induces oscillons from the fluctuations of the untwisted fields during collisions. Twisted fields have negative energies ~\cite{FordToms}. Correspondingly, structures emerging from them, antioscillons, can have negative energies.

\subsection{Setup}
Consider the theory of a scalar field with potential $V(\phi, T)$, with false vacuum $\phi_F = 0$ and true vacuum $\phi = \phi_T$.
At $T=0$, the nucleation rate $\Gamma$ is related to the Euclidean 4-action $S_4[\phi_{bounce}]$ by $\Gamma = Ae^{-S_4[\phi_{bounce}]}$.
$A$ is a prefactor describing the effect of small fluctuations around the bounce solution.
Normally $\Gamma$ is small in the dilute gas, thin wall approximation of bubbles.
I work within the context of these approximations here.

It should be noted that in the cosmological context, the dilute gas of bubbles is in a bath at $T \neq 0$.
As pointed out by Linde~\cite{ref:Linde}, the quantum statistics of the dilute gas of bubbles at temperature $T$ in $(D-1)$-dimensions is formally equivalent to quantum field theory in $D$-dimensions with the ``time'' direction periodic with periodicity $\beta = T^{-1}$.  The action in this case becomes $S_4[\phi] = \beta S_3[\phi] = T^{-1}\int d^3x \left[ \frac{1}{2}(\nabla\phi)^2 + V(\phi,T) \right]$, and the dominant contribution to the Euclidean path integral corresponds to cylindrical $O(3)$ bounce solutions, which have $O(3)$ symmetry in the spatial directions and are periodic in the Euclidean time direction $\tau$ with period $\beta$.
The nucleation rate of these cylindrical solutions is given by $\Gamma = AT^4e^{-S_3[\phi_{bounce}]/T}$ where $A$ is again related to the two-point function of fluctuations. Generally, the type of effective potential that gives rise to false vacuum decay can be expressed as
\begin{equation}
  V(\phi, T) = \frac{1}{2}m^2(T)\Phi^2 - \frac{1}{3}\alpha(T)\Phi^3 + \frac{\lambda(T)}{4}\Phi^4
  \label{eqn:pottemp}
\end{equation}
Let us for the moment focus on the familiar untwisted field populating a bubble.
Linear fluctuations of this field around the vacuum in k-space satisfy
\begin{equation}
  \ddot{\delta\phi} + \left(k^2 + V''(\phi_F, T)\right)\delta\phi = 0 \,
  \label{eqn:linfluc}
\end{equation}
Oscillons can emerge in the vicinity of collisions as resonant nucleation out of these fluctuations when bubble collisions kick the field, e.g. from $\phi \approx \phi_0 + \delta\phi$ where $\phi_0$ is the zero-mode oscillation around the vacuum state. Since collisons are directional, say along the $z$-axis, and break the $O(3)$ symmetry, then the gradient term above is replaced by $k^2 = (k_{x}^2 + k_{y}^2 - \nabla_{z}^{2}) $. With this reminder, we can see from~\eqref{eqn:linfluc} that the field $\delta\phi$ becomes unstable and can grow exponentially when $V'' < 0$ for certain values of $k^2$.
Physically, the large fluctuation in the field is transferred to higher $k$-modes due to nonlinear scattering and some of the $k < V''$ are amplified. Therefore, the homogeneity in the bubbles is broken during the collision and this breaking is adequately described by the cylindrical $O(3)$ symmetry of a single bubble going to $O(2)$ symmetry of the 2 colliding bubble system.
The oscillon structures emerging from this process are roughly given by $\phi_{osc}(t,r) \approx \phi_o(t)e^{-r^2/R^2}$ where $R^{-1}$ is the upper limit in the amplified k-band, $0< k < 2R^{-1}$.

But due to the nontrivial topology, the bubbles admits the twisted configuration also ~\cite{ref:FordToms}, $\psi(x,T)$, with a potential similar to ~\eqref{eqn:pottemp}) and antiperiodic boundary conditions (in the $\tau$ direction).
Later we allow for coupling between the two fields of the form $H_{int} = g^2\phi^2\psi^2$. The 2-point function for the free fields, calculated in~\cite{ref:FordToms} is given by the periodicity length $\beta$,  $\langle\phi_{\beta}^2\rangle = \frac{12\pi^2}{\beta^2}$ and $\langle\psi_{\beta}^2\rangle = \frac{-24\pi^2}{\beta^2}$

\section{Antioscillons}
 

Here we ignore coupling between the two fields. The action of cylindrical bubbles is $S[\phi,\psi,T] = S_3/T$, where $S_3[\phi,\psi,T] = -\frac{4}{3}\pi\mathcal{r}^3\epsilon + 4\pi\mathcal{r}^2\sigma(T)$,
and $\epsilon = V(\phi_{true}) - V(\phi_{false})$ with $\sigma(T)$ the surface tension of the bubble. For our potential~\eqref{eqn:pottemp}, $\sigma = \alpha^2 / (12\lambda^2)$, and the energy of the twisted field is $\langle\psi^2\rangle_{\beta} = -24\pi^2/\beta^2$ or $ \langle\psi^2\rangle_{\beta} = -24\pi^2T_{dS}^2$
where $T_{dS}$ is the temperature of the bath of the embedding deSitter background, $T_{dS} = H_{dS}/2\pi = \sqrt{\epsilon}/2\pi$. The energy in this field is less than that of the bubble tension $\sigma$.
Although $\psi$ has negative energy, it is completely harmless and the system of the bubble and the field is stable.
Bubble collisions 'kick' this field up the potential, and scattering excites the k-modes in the instability band, in complete analogy with the untwisted fields, namely $\ddot{\delta\psi} + (k_{x}^2 +k^{2}_{y} - \nabla_{z}^2 + V''(\psi,T))\delta\psi = 0$. For a sufficiently strong collision, the field can be driven to fluctuate to the region $V''<0$ where the resonant condensation of fluctuations into antioscillons for certain $k$ modes ( modes that make the term in brackets negative) occurs. Condensation of the amplified fluctuations into localized energy lumps becomes the antioscillons. Ultimately the antioscillons decay and the bubble is not unstable.

\subsection{Interactions and Instabilities}
The untwisted field $\phi$ is normally present alongside the twisted fields $\psi$ in the bubble. Couplings of the form $H_{int}=g^2\phi^2\psi^2$ can occur and will show up in 
the potential $V(\phi,\psi,T) = \frac{m_{\phi}^2}{2}\phi^2 - \frac{\alpha_{\phi}}{3}\phi^3 + \frac{\lambda_{\phi}}{4}\phi^4 + \frac{m_{\psi}^2}{2}\psi^2 - \frac{\alpha_{\psi}}{3}\psi^3 + \frac{\lambda_{\psi}}{4}\psi^4 $.
Such couplings are problematic for the stability of the untwisted field ~\cite{ref:FordToms} since they can assign an effective tachyonic mass to the untwisted field.
Let us imagine that $\hat{\psi}$ is in a coherent state.
The field equation for $\phi$ with mass $m_{\phi}$ is $\Box\phi + V'(\phi,T) + g^2\langle\psi^2\rangle_{\beta}\phi = 0$. If for the moment we approximate $V'(\phi,T) \approx m_{\phi}^2(T)\phi$, then from the last equation we see that instability occurs anytime that $g^2\langle\psi^2\rangle_{\beta} < -m_{\phi}^2(T)$. The instability in the interior of these bubbles can grow without bound. With both fields present on the nontrivial toplogical background of the bubble, their interaction leads to two scenarios:
(i) if the instability grows without bound after the bubble nucleation, then the growth of the instability backreacts on the bubbles and they may vanish. The field dynamics is very complicated, but depending on the timescale $\mu^{-1}$ for the instability, the phase transition may be incomplete and the field pushed back to the false vacuum;
(ii) if the interaction and subsequent growth of modes occurs in the false vacuum before bubbles nucleate, then the resulting quantum instability of the two fields can result in a classical cross-over the barrier by thermal instead of quantum fluctuations.  In this case the phase transition is completed almost instantaneously.
\subsection{Conclusions}
I have described how the nontrivial topology of bounce solutions at finite temperature can result in a new field configuration with negative energy \cite{lauraoscillon}.
Like the familiar untwisted field configuration, during bubble collisions the twisted field can be driven to large fluctuations, where a band of high k-modes is parametrically amplified.
This results in the formation of new long-lived structures, antioscillons. Instability from the two interacting field influences the completion of the phase transition.
The growth of modes can destroy the bubble or large quantum instabilities can drive false vacuum decay via a thermal crossover the barrier instead of tunneling.
Backreaction of these modes on the effective potential becomes important.


\end{document}